\documentclass[10pt,twocolumns]{IEEEtran}

\usepackage{times,graphics,amssymb,amsmath,bm,cite}
\usepackage[dvips]{graphicx}
\usepackage[english]{babel}
\usepackage{latexcad}
\usepackage{epsfig,psfig,latexsym}

\newcommand{\eeq}{\end{equation}}

\newcommand{\bJ}{\mbox{\boldmath $J$}}

\newcommand{\bM}{\mbox{\boldmath $M$}}

\newcommand{\bh}{\mbox{\boldmath $h$}}
\newcommand{\bH}{\mbox{\boldmath $H$}}

\newcommand{\bx}{\mbox{\boldmath $x$}}

\newcommand{\bn}{\mbox{\boldmath $n$}}

\newcommand{\bd}{\mbox{\boldmath $d$}}
\newcommand{\bQ}{\mbox{\boldmath $Q$}}
\newcommand{\bO}{\mbox{\boldmath $O$}}

\newcommand{\by}{\mbox{\boldmath $y$}}
\newcommand{\ds}{\displaystyle}

\newcommand{\K}{{\cal K}}
\newcommand{\G}{{\cal G}}
\newcommand{\N}{{\cal N}_0}
\def\R{{\cal R}}
\def\S{{\cal S}}

\newcommand{\beq}{\begin{equation}}

\def\QED{\mbox{\rule[0pt]{1.5ex}{1.5ex}}}

\begin{document}


\thispagestyle{empty}
\title{Power Control and Receiver Design for Energy Efficiency in
Multipath CDMA Channels with Bandlimited Waveforms}
\author{{Stefano Buzzi$^1$, Valeria Massaro$^1$ and H. Vincent Poor$^2$\thanks{The research was supported in part by the U. S. Air Force Research Laboratory under Cooperative Agreement No.
FA8750-06-1-0252 and by the U. S. Defense Advanced Research
Projects Agency under Grant HR0011-06-1-0052.
}}\\
{\normalsize $^1$Universit\`{a} degli Studi di Cassino, Cassino (FR), ITALY 03043}\\
{\normalsize  \{buzzi, massaro\}@unicas.it} \\
{\normalsize $^2$School of Engineering and Applied Science -- Princeton University} \\
{\normalsize Princeton, NJ, 08544, USA; poor@princeton.edu}
}

 \maketitle
\date{January 3, 2007}

\vspace*{-.3cm}
\begin{abstract}
This paper is focused on the cross-layer design problem of joint
multiuser detection and power control for  energy-efficiency
optimization in a wireless data network through a game-theoretic
approach. Building on work of Meshkati, et al., wherein the tools
of game-theory are used in order to achieve energy-efficiency in a
simple synchronous code division multiple access system, system
asynchronism, the use of bandlimited chip-pulses, and the
multipath distortion induced by the wireless channel are
explicitly incorporated into the analysis. Several non-cooperative
games are proposed wherein users may vary their transmit power and
their uplink receiver in order to maximize their utility, which is
defined here as the ratio of data throughput to transmit power. In
particular, the case in which a linear multiuser detector is
adopted at the receiver is considered first, and then,  the more
challenging case in which non-linear decision feedback multiuser
detectors are employed is considered. The proposed games are shown
to admit a unique Nash equilibrium point, while simulation results
show the effectiveness of the proposed solutions, as well as that
the use of a decision-feedback multiuser receiver brings
remarkable performance improvements.

\end{abstract}


\section{Introduction} \label{sec:introduction}
Game theory \cite{gtbook} is a branch of mathematics that has been
applied primarily to social science and economics to study the
interactions among several autonomous subjects with contrasting
interests. Recently, it has been discovered that it can also be
used for the design and analysis of communication systems, mostly
with application to resource allocation algorithms \cite{gt}, and,
in particular, to power control \cite{yates}. As examples, the
reader is referred to \cite{nara1,nara2}. Here, for a multiple
access wireless data network, noncooperative and cooperative games
are introduced, wherein users choose their transmit powers in order
to maximize their own utilities, defined as the ratio of the
throughput to transmit power. While the above studies consider the
issue of power control assuming that a conventional matched filter
is available at the receiver, the recent paper \cite{meshkati}
considers for the first time the problem of joint linear receiver
design and power control so as to maximize the utility of each
user. In particular, it is shown here that the inclusion of
receiver design in
the considered game brings remarkable advantages, and, also,
results based on the powerful large-system analysis are presented.

All of the cited studies, while laying the foundations of the
game-theoretic approach to utility maximization in wireless data
networks, focus on a very simple model, i.e. a synchronous direct
sequence code division multiple access (DS/CDMA) channel subject
to flat-fading. In this paper, instead, we extend the
game-theoretic framework to a more practical and challenging
scenario, namely we explicitly take into account (a) the possible
system asynchrony across users; (b) the use of bandlimited
chip-pulses; and (c) the multipath distortion induced by the
wireless propagation channel. Note that in such a scenario
intersymbol and interchip interference arises, thus implying that
the appealing mathematical relationships between the
signal-to-interference plus noise ratio (SINR) and the transmit
power (as revealed in \cite{meshkati}) do not hold any longer, and
this makes system analysis much more involved than it is for the
case in which no self-interference exists. A further contribution
of this paper is the consideration of non-linear multiuser
receivers. Indeed, while previous studies have considered the case
in which either a matched filter (see, e.g., \cite{nara2}) or a
linear multiuser detector \cite{meshkati} is adopted at the uplink
receiver, here we also consider the case in which a non-linear
decision feedback receiver is employed at the receiver.\\
\noindent {\bf Notation:} $(\cdot)^T$ denotes
transpose, while $\ast$ and
$\times$ denote linear convolution and ordinary product,
respectively.


\section{Preliminaries and Problem Statement}

Consider the uplink of an  asynchronous DS/CDMA system with $K$
users, employing bandlimited chip pulses and operating over a
frequency-selective fading channel. The received signal at the
access point (AP) may be  written as\footnote{For the sake of
simplicity we assume here a real signal model; however, the
extension to complex signals to account for I and Q components is
trivial.} \beq r(t)=\ds \sum_{p=0}^{B-1} \sum_{k=1}^{K} \sqrt{p_k}
b_k(p) s'_{k}(t -\tau_k -pT_b) * c_k(t) + w(t) \, .
\label{eq:r(t)} \eeq In the above expression, $B$ is the
transmitted frame or packet length, $T_b$ is the bit-interval
duration, $p_k$ and  $\tau_k \geq 0$ denote the transmit power and
timing offset of the $k$-th user, $b_k(p) \in \{+1,-1\}$ is the
$k$-th user's information symbol in the $p$-th signaling interval
(extension to modulations with a larger cardinality is
straightforward). Moreover, $c_k(t)$ is the impulse response
modeling the channel effects between the receiver and the $k$-th
user's transmitter, while $w(t)$ is the additive noise term, which
is assumed to be a zero-mean, Wide-Sense Stationary (WSS)
 white Gaussian process with Power Spectral Density (PSD)
$\N/2$. It is also assumed that the channel coherence time exceeds
the packet duration $BT_b$, so that the channel impulse responses
$c_0(t), \ldots, c_{K-1}(t)$ may be assumed to be time-invariant
over each transmitted frame. As to $s'_{k}(t)$, it is the $k$-th
user's signature waveform and is written as
$
s'_{k}(t)=\ds \sum_{n=0}^{N-1} \beta_{k}^{(n)} h_{\rm
SRRC}(t-nT_c),
$
with $\{\beta_{k}^{(n)}\}_{n=0}^{N-1}$ the $k$-th user's spreading
sequence, $N$ the processing gain, $T_c=T_b/N$ the chip interval,
and $h_{\rm SRRC}(\cdot)$ a square-root raised cosine waveform
with roll-off factor $\alpha \in [0,1]$.
We assume here that $h_{\rm SRRC}(t)$ is zero outside the interval $[0,
4T_c]$ and attains its maximum value in $t=2T_c$.

The receiver front-end consists of a filter with impulse response
$h_{\rm SRRC}(-t)$, followed by a sampler at rate $M/T_c$;
in our simulations we will assume that $M=2$. Denoting by $y(t)$
the signal at the output of the receiver matched filter, it can be
easily shown that \beq
\begin{array}{cc}
y(t)= r(t) * h_{\rm SRRC}(-t)=
\ds \sum_{p=0}^{B-1} \sum_{k=1}^{K} \sqrt{p_k} b_k(p)  \times \\
s_{k}(t -\tau_k -pT_b) * c_k(t) + \underbrace{w(t)* h_{\rm
SRRC}(-t)}_{n(t)} \, , \label{eq:y(t)}
\end{array}
\eeq with
$
s_{k}(t)=s'_{k}(t) * h_{\rm SRRC}(-t) =\ds \sum_{n=0}^{N-1}
\beta_{k}^{(n)} h_{\rm RC}(t-nT_c),
$
and $h_{\rm RC}(t)=h_{\rm SRRC}(t) * h_{\rm SRRC}(-t)$.

Denoting by $h_{k}(t)=s_{k}(t -\tau_k) * c_k(t)$ the {\em
effective} signature waveform for the $k$-th user in the $p$-th
signaling interval, the signal (\ref{eq:y(t)}) can be expressed as
\beq y(t)=\ds \sum_{p=0}^{B-1} \sum_{k=1}^{K} \sqrt{p_k} b_k(p)
h_{k}(t -pT_b)  + n(t) \; . \label{eq:r(t)2} \eeq Notice that the
waveform $h_{k}(t)$ is supported in the interval $[\tau_k, \tau_k
+T_b +T_m+7T_c]$, where $T_m$ denotes the maximum channel
multipath delay spread over the $K$ active users.
Assuming that $\tau_k +T_m< T_b$, the support of the waveform
$h_{k}(t-pT_b)$ is contained in the interval $[pT_b,
(p+2)T_b+7T_c]$, thus implying that, for a system with processing
gain larger than 7,  in the symbol interval ${\cal
I}_p=[pT_b,(p+2)T_b]$ the contribution from at most four symbols for
each user (i.e. the $p$-th, the $(p-1)$-th, the $(p-2)$-th and the
$(p+1)$-th ones) is observed. Accordingly, sampling the waveform
$y(t)$ at rate $M/T_c$, the $2MN$-dimensional vector $\by(p)$
collecting the data samples of the interval ${\cal I}_p$ can be
expressed as \beq
\begin{array}{llll}
\by(p)=&\ds \sum_{k=1}^{K} \sqrt{p_k} \left[b_k(p-2) \bh_{k,-2} + b_k(p-1) \bh_{k,-1} + \right. \\ & \left.
b_k(p) \bh_{k,0} + b_k(p+1) \bh_{k,+1}\right] + \bn(p)\; .
\end{array}
\label{eq:discretedata} \eeq In (\ref{eq:discretedata}), the
vector $\bh_{k,i}$ is $2MN$-dimensional, and contains the samples
of the signature $h_{k}(t-(p+i)T_b)$ coming from ${\cal I}_p$,
while the vector $\bn(p)$ contains the noise contribution, and is
a Gaussian random vector with covariance matrix $\bM$. We assume
that the data vector $\by(p)$ will be used in order to detect the
information symbols $b_1(p), b_2(p), \ldots, b_{K}(p)$, i.e. the
$p$-th epoch data symbols for all the users.

Assume now that each mobile terminal is interested both in having
its data received with as small as possible error probability at
the AP, and in making optimal use of the energy stored in its
battery. Obviously, these are conflicting goals, since error-free
reception may be achieved by increasing the received SNR, i.e. by
increasing the transmit power, which of course comes at the
expense of battery life\footnote{Of course there are many other
strategies to lower the data error probability, such as for
example the use of error correcting codes, diversity exploitation,
and implementation of optimal reception techniques at the
receiver. Here, however, we are mainly interested in energy
efficient data transmission and power usage, so we assume that
only the transmit power and the receiver strategy can be varied to
achieve energy efficiency.}. A useful approach to quantify these
conflicting goals is to define the utility of the $k$-th user as
the ratio of its throughput, defined as the number of information
bits that are received with no error in the unit time, to its
transmit power \cite{nara1,nara2}, i.e. \beq u_k=\ds
\frac{T_k}{p_k}\; . \label{eq:utility} \eeq Note that $u_k$ is
measured in bits/Joule. Denoting by $R$ the common rate of the
network and assuming that each packet of $B$ bits contains $L$
information bits and $B-L$ overhead bits, reserved, e.g., for
channel estimation and/or parity checks, the throughput $T_k$ can
be expressed as \beq T_k=\ds R \frac{L}{B} P_k \label{eq:Tk} \eeq
wherein $P_k$ denotes the probability that a packet from the
$k$-th user is received error-free. In the considered DS/CDMA
setting, the term $P_k$ depends formally on a number of parameters
such as the spreading codes of all the users,  their  transmit
powers and their channel impulse responses; however, a customary
approach is to model the overall interference  as a Gaussian
random process, and assume that $P_k$ is an increasing function of
the $k$-th user's SINR $\gamma_k$, which is a good model for many
practical scenarios.

For the case in which a linear receiver is used to detect the data
symbol $b_k(p)$, according, i.e., to the decision rule \beq
\widehat{b}_k(p)=\mbox{sign}\left[\bd_k^T \by(p)\right] \; ,
\label{eq:decrule} \eeq with $\widehat{b}_k(p)$ the estimate of
$b_k(p)$ and $\bd_k$ the $2NM$-dimensional vector representing the
receive filter for user $k$, it is easily seen that for the case
at hand the SINR $\gamma_k$ can be written as \beq \gamma_k=\ds
\frac{p_k (\bd_k^T \bh_{k,0})^2}{\bd_k^T \bM \bd_k + \ds \sum_{i
\neq k} \sum_{j=-2}^1 p_i (\bd_k^T \bh_{i,j})^2 + \sum_{j \neq 0}
p_k (\bd_k^T \bh_{k,j})^2} \; . \label{eq:gamma} \eeq


The exact shape of $P_k(\gamma_k)$ depends not only on $\gamma_k$,
but also on other factors such as the modulation and coding type.
However, in all cases of relevant interest, it is an increasing
function of $\gamma_k$ with a sigmoidal shape, and converges to
unity as $\gamma_k \rightarrow + \infty$; as an example, for
binary phase-shift-keying (BPSK) modulation coupled with no
channel coding, it is easily shown that \beq
P_k(\gamma_k)=\left[1-Q(\sqrt{2\gamma_k})\right]^B \; ,
\label{eq:psr} \eeq with $Q(\cdot)$ the complementary cumulative
distribution function of a zero-mean random Gaussian variate with
unit variance.

It should be noted however that substituting Eq. (\ref{eq:psr})
into (\ref{eq:Tk}), and, in turn, into (\ref{eq:utility}), leads
to a strong incongruence. Indeed, for $p_k \rightarrow 0$, we have
$\gamma_k \rightarrow 0$, {\em but} $P_k$ converges to a small but
non-zero value (i.e. $2^{-B}$), thus implying that an unboundedly
large utility can be achieved by transmitting with zero power. To
circumvent this problem, a customary approach
\cite{nara2,meshkati} is to replace $P_k$ with an {\em efficiency
function}, say $f_k(\gamma_k)$, whose behavior should approximate
as close as possible that of $P_k$, except that for $\gamma_k
\rightarrow 0$  it is required that $f_k(\gamma_k)= o(\gamma_k)$.
The function $f(\gamma_k)=(1-e^{-\gamma_k/2})^B$ is a widely
accepted substitute for the true probability of correct packet
reception, and in the following we will adopt this model. This
efficiency function is increasing and S-shaped, converges to unity
as $\gamma_k$ approaches infinity, and has a continuous first
order derivative\footnote{Note that we have omitted the subscript
\emph{k}, i.e. we have used the notation $f(\gamma_k)$ in place of
$f_k(\gamma_k)$ since we assume that the efficiency function is
the same for all the users.}.

Summing up, substituting (\ref{eq:Tk}) into (\ref{eq:utility}) and
replacing the probability $P_k$ with the above defined efficiency
function, we obtain the following expression for the $k$-th user's
utility: \beq u_k=R \ds \frac{L}{B} \frac{f(\gamma_k)}{p_k} \; ,
\quad \forall k=1, \ldots, K \; . \label{eq:utility2} \eeq

\section{Non-cooperative games with linear receivers}

In what follows we illustrate three different noncooperative games
wherein each user aims at maximizing its own utility by varying
its transmit power, and, possibly, its linear uplink receiver.
Formally, the considered game $\G$ can be described as the triplet
$\G=\left[\K, \left\{\S_k\right\}, \left\{u_k\right\} \right]$,
wherein $\K= \left\{1, 2, \ldots, K\right\}$ is the set of active
users participating in the game, $u_k$ is the $k$-th user's
utility defined in eq. (\ref{eq:utility2}), and \beq \S_k=[0,
P_{k,\max}] \times \R^{2NM} \; , \label{eq:strategy} \eeq is the
set of possible actions (strategies) that user $k$ can take. It is
seen that $\S_k$ is written as the Cartesian product of two
different sets, and indeed $[0, P_{k, \max}]$ is the range of
available transmit powers for the $k$-th user (note that $P_{k,
\max}$ is the maximum allowed transmit power of user $k$), while
$\R^{2NM}$, with $\R$ the real line, defines the set of all
possible linear receive filters.

\subsection{Power control with plain matched filter}
We first consider the case in which $\S_k=[0, P_{k,\max}]$ and the
uplink receiver is a matched filter, i.e. we assume that each user
tunes its transmit power in order to maximize its own utility, but
the uplink receiver is a matched filter\footnote{Note that for an
oversampling factor $M>1$ a whitening transformation would in
principle be required prior to matched filtering; for the sake of
simplicity, however, noise whitening is not performed here.}.
Consequently, the $k$-th user's SINR is expressed as \beq
\gamma_k\!=\ds \frac{p_k \|\bh_{k,0}\|^4}{\bh_{k,0}^T \bM
\bh_{k,0} \!+ \!\ds \sum_{i \neq k} \!\sum_{j=-2}^1 \!p_i
(\bh_{k,0}^T \bh_{i,j})^2 \!+\! \sum_{j \neq 0} p_k (\bh_{k,0}^T
\bh_{k,j})^2} \; , \label{eq:gammamf} \eeq and the noncooperative
game can be cast as the following maximization problem \beq \ds
\max_{\S_k} u_k = \!\!\!\max_{p_k \in [0, P_{k,\max}]}
\!u_k(p_k)\! = \!\ds\! \max_{p_k \in [0, P_{k,\max}]}
\!\frac{f(\gamma_k(p_k))}{p_k}\; , \label{eq:game} \eeq $\forall
k=1, \ldots, K$. Now, the following result can be stated about
maximization (\ref{eq:game}).

\noindent {\bf Proposition 1:} {\em The non-cooperative game
defined in (\ref{eq:game}) admits a unique Nash equilibrium point
$p_k^*$, for $k=1, \ldots, K$, wherein $p_k^*=\min \{\bar{p}_k,
P_{k, \max} \}$, with $\bar{p}_k$ denoting the $k$-th user's
transmit power such that the $k$-th user's  SINR
$\gamma_k$ equals $\bar{\gamma}_k$, i.e. the unique solution of
the equation \beq \ds \frac{B}{2a_k}\gamma (a_k -b_k \gamma)=
\exp(\gamma/2)-1 \, , \label{eq:nuova} \eeq with
$a_k=\|\bh_{k,0}\|^4$ and
$$ b_k=\sum_{j
\neq 0}(\bh_{k,0}^T \bh_{k,j})^2  .$$ } \noindent {\bf Proof:} The
proof is omitted for brevity. \hfill \rule{2.5mm}{2.5mm}

In summary, Proposition 1 states that a Nash equilibrium for the
noncooperative game (\ref{eq:game}) always exists, and it can be
found with the following steps. First, the unique solution
$\bar{\gamma}_k$ of the equation (\ref{eq:nuova}) is determined.
Then, each user adjusts its transmit power to achieve its
target SINR $\bar{\gamma}_k$. These steps are repeated until
convergence is reached.

\subsection{Power control and receiver design with no ISI}
Let us now consider the case in which not only the transmit power,
but also the linear receiver can be tuned so as to maximize
utility for each user; moreover, let us also impose the condition
that the receive filter be orthogonal to the subspace spanned by
ISI. Denoting by $\bO_k$ a $2NM \times (2NM-3)$-dimensional matrix
containing in its columns a basis for the orthogonal complement of
the subspace spanned by the $k$-th user's ISI, i.e. by the vectors
$\bh_{k,-2}$, $\bh_{k,-1}$, and $\bh_{k,1}$, we assume that the
decision rule to detect the symbol $b_k(p)$ can be written as \beq
\widehat{b}_k(p)=\mbox{sign}\left[\bx_k^T \bO_k^T \by(p)\right] \;
, \label{eq:decrule2} \eeq with $\bx_k$ a $(2NM-3)$-dimensional
vector. The $k$-th user's SINR is now written as \beq \gamma_k=\ds
\frac{p_k (\bx_k^T \bO_k^T \bh_{k,0})^2}{\bx_k^T \bO_k^T  \bM
\bO_k \bx_k + \ds \sum_{i \neq k} \sum_{j=-2}^1 p_i (\bx_k^T
\bO_k^T  \bh_{i,j})^2} \; , \label{eq:gamma2} \eeq namely the
$k$-th user's transmit power appears only in the numerator in the
RHS of (\ref{eq:gamma2}), thus implying that the relation $
\frac{d \gamma_k}{dp_k}=
 \frac {\gamma_k}{p_k}$ holds. We now consider the
following maximization problem \beq \ds \max_{\S_k} u_k =
\max_{p_k, \bx_k} u_k(p_k, \bx_k ) \; , \quad \forall k=1, \ldots,
K \; . \label{eq:game2} \eeq Given (\ref{eq:utility2}), the above
maximization can be also written as  \beq \ds \max_{p_k, \bx_k}
\frac{f(\gamma_k(p_k, \bx_k))}{p_k}= \max_{p_k} \frac{f\left(\ds
\max_{\bx_k}\gamma_k(p_k, \bx_k)\right)}{p_k} \; ,
\label{eq:deriv} \eeq i.e. we can first take care of SINR
maximization with respect to  linear receivers, and then focus on
maximization of the resulting utility with respect to transmit
power. We now have the following:

\noindent {\bf Proposition 2:} {\em Let $\bM_{\by \by}$ denote the
covariance matrix of the vector $\by(p)$. The non-cooperative game
defined in (\ref{eq:game2}) admits a unique Nash equilibrium point
$(p_k^*, \bx_k^*)$, for $k=1, \ldots, K$, wherein
\begin{itemize}
\item[-]
$\bx^*_k=\sqrt{p_k} \left( \bO_k^T \bM_{\by \by}
\bO_k\right)^{-1}\bO_k^T\bh_{k,0}$ is the unique (up to a positive
scaling factor) $k$-th user's receive filter that maximizes the  SINR
$\gamma_k$ in (\ref{eq:gamma2}). Denote
$\gamma_k^*=\max_{\bx_k}\gamma_k$.
\item[-]
$p_k^*=\min \{\bar{p}_k, P_{k, \max} \}$, with $\bar{p}_k$ the
$k$-th user's transmit power such that the $k$-th user's maximum
SINR $\gamma_k^*$ equals $\bar{\gamma}$, i.e. the unique solution
of the equation $f(\gamma)=\gamma f'(\gamma)$, with $f'(\gamma)$
denoting the derivative of $f(\gamma)$.
\end{itemize}}
\noindent {\bf Proof:} The proof is omitted due to lack of space.
Note however that, due to the constraint that the receive filter
is orthogonal to the ISI contribution, the mathematical structure
of the maximization (\ref{eq:game2}) is similar to that of the
noncooperative game proposed in \cite{meshkati}, and the proof can
thus be adapted from there. \hfill \rule{2.5mm}{2.5mm}

The above equilibrium can be reached according to the following
procedure. For a given set of users' transmit powers, the receiver
filter coefficients can be set according to the relation
$\bx^*_k=\sqrt{p_k} \left( \bO_k^T \bM_{\by \by}
\bO_k\right)^{-1}\bO_k^T\bh_{k,0}$; each user can then tune its
power so as to achieve the target SINR $\bar{\gamma}$. These steps
are repeated until convergence is reached.

\subsection{Power control and unconstrained receiver design}
Finally, we consider the case in which no constraint is imposed on
the receive filter, so that the $k$-th user's SINR is written as
in Eq. (\ref{eq:gamma}). We now consider the following
maximization \beq \ds \max_{\S_k} u_k = \max_{p_k, \bd_k} u_k(p_k,
\bd_k )= \max_{p_k}  \frac{f\left(\ds \max_{\bd_k}\gamma_k(p_k,
\bd_k)\right)}{p_k} \, , \label{eq:game3} \eeq $ \forall k=1,
\ldots, K $, wherein the fact that the efficiency function is
non-decreasing has been exploited. Now, the maximization of
$\gamma_k$ with respect to $\bd_k$ is trivial, since it is well
known that the linear receiver that maximizes SINR is the minimum
mean square error multiuser receiver. As a consequence, denoting
by $\bar{\bd}_k$ the maximizer of $\gamma_k$, we have \beq
\bar{\bd}_k=\sqrt{p_k} \bM_{\by \by}^{-1} \bh_{k,0} \, ;
\label{eq:mmsereceiver} \eeq let us denote by
$\bar{\gamma}_k(p_k)$ the $k$-th user's SINR with
$\bd_k=\bar{\bd}_k$. Maximizing the utility with respect to the
transmit power requires instead solving the equation \beq
f(\bar{\gamma}_k(p_k))= f'(\bar{\gamma}_k(p_k))
\bar{\gamma}_k'(p_k) p_k \; , \label{eq:casino} \eeq with
$(\cdot)'$ denoting first-order derivative with respect to $p_k$.
Now,  (\ref{eq:casino}) appears to be quite complicated and
unmanageable. Indeed, note that letting $\bH_k=[\bh_{k,-2} \;
\bh_{k,-1} \; \bh_{k,1}]$, we have \beq \bM_{\by \by}= \bM_{\by
\by}(p_k)= \bQ_k + p_k \bH_k\bH_k^T + p_k \bh_{k,0}  \bh_{k,0}^T
\; , \eeq with $\bQ_k$ the covariance matrix of the thermal noise
and of the multiuser interference for the $k$-th user, thus
implying that $\bar{\gamma}_k(p_k)$ is expressed as \beq
\bar{\gamma}_k(p_k)=\ds \frac {p_k (\bh_{k,0}^T \bM_{\by
\by}^{-1}(p_k)\bh_{k,0})^2}{\bh_{k,0}^T \bM_{\by \by}^{-1}(p_k)
\bh_{k,0} - p_k (\bh_{k,0}^T \bM_{\by \by}^{-1}(p_k)\bh_{k,0})^2}
\; . \label{eq:gammacasino} \eeq It is clear that substituting
(\ref{eq:gammacasino}) and its first-order derivative into
(\ref{eq:casino}) and solving with respect to $p_k$ is quite
complicated. Accordingly, we have not been able in this case to
formally prove the existence of a Nash equilibrium point. However,
we have numerically evaluated the utility function and
(\ref{eq:casino}), and we have found in every case considered that
(\ref{eq:casino}) admits a unique solution and that the resulting
game admits an equilibrium point. We thus state the following
conjecture.

\noindent {\bf Conjecture 1:} {\em The non-cooperative game
defined in (\ref{eq:game3}) admits a unique Nash equilibrium point
$(p_k^*, {\bd}^*_k)$, for $k=1, \ldots, K$, wherein
\begin{itemize}
\item[-]
$\bd^*_k$ is the linear MMSE receiver (see Eq.
(\ref{eq:mmsereceiver})), which maximizes the SINR $\gamma_k$ in
(\ref{eq:gamma}). Denote $\bar{\gamma}_k=\max_{\bd_k}\gamma_k$.
\item[-]
$p_k^*=\min \{\bar{p}_k, P_{k, \max} \}$, with $\bar{p}_k$ the
unique solution of Eq. (\ref{eq:casino}).
\end{itemize}}
\noindent Also in this case, the equilibrium can be reached
through an iterative procedure. For a given set of users' transmit
powers, the receiver filter coefficients can be set equal to the
MMSE multiuser receiver; each user can then tune its power to
$p_k^*$, and these steps are repeated until convergence is
reached.

\section{Non-cooperative games with decision-feedback receivers}
Consider now the case in which a non-linear decision feedback receiver
is used at the receiver.  We assume that the users are indexed
according to a non-increasing sorting of their channel gains, i.e.
we assume that $\|\bh_{1,0}\| > \|\bh_{2,0}\| > \ldots, \|
\bh_{K,0}\|$. We consider a serial interference cancellation (SIC)
receiver wherein detection of the symbol from the $k$-th user is made
according to the following rule \beq
\widehat{b}_k(p)=\mbox{sign}\left[\bd_k^T \left(\by(p) -
\sum_{j<k} \sum_{i=-2}^0 \sqrt{p_j}  \widehat{b}_j(p+i) \bh_{j,i}
\right) \right] \; . \label{eq:decruleSIC} \eeq
Accordingly, if past decisions are correct, users that are detected later
enjoy a considerable reduction of multiple access interference,
and indeed the SINR for user $k$, under the assumption of
correcteness of past decisions, is written as
\beq \gamma_k=\ds
\frac{p_k  (\bd_k^T \bh_{k,0})^2}{\zeta_k} \; , \label{eq:gammaSIC} \eeq
with $\zeta_k=\bd_k^T \bM \bd_k + \ds \sum_{j
< k} p_j (\bd_k^T \bh_{j,1})^2 + \ds \sum_{j \neq 0} p_k (\bd_k^T
\bh_{k,j})^2 + \ds \sum_{j > k} \sum_{i=-2}^{1}p_j (\bd_k^T
\bh_{j,i})^2$.

\subsection{Power control and receiver design with no ISI}
Replicating the path of the previous section,  we start imposing
the constraint that the receive filter be orthogonal to the ISI
subspace for each user, i.e. our decision rule is \beq
\widehat{b}_k(p)=\mbox{sign}\left[\bx_k^T \bO_k^T \left(\by(p) -
\sum_{j<k} \sum_{i=-2}^0 \sqrt{p_j}  \widehat{b}_j(p+i) \bh_{j,i}
\right) \right] \; , \label{eq:decruleSIC2} \eeq and the $k$-th
user SINR is \beq \gamma_k=\ds \frac{p_k  (\bx_k^T
\bO_k^T\bh_{k,0})^2}{\varrho_k} \; , \label{eq:gammaSIC2} \eeq
with $\varrho_k=\bx_k^T \bO_k^T \bM \bO_k \bx_k + \ds \sum_{j < k}
p_j (\bx_k^T \bO_k^T \bh_{j,1})^2  + \ds \sum_{j > k}
\sum_{i=-2}^{1}p_j (\bx_k^T \bO_k^T \bh_{j,i})^2$. Given receiver
(\ref{eq:decruleSIC2}) and the SINR expression
(\ref{eq:gammaSIC2}), we consider here the problem of utility
maximization with respect to the transmit power,  and receiver
vectors $\bx_1, \ldots, \bx_K$: \beq \ds \max_{p_k, \bx_k}
\frac{f(\gamma_k(p_k, \bx_k))}{p_k} \; , \quad \forall k=1,
\ldots, K \; . \label{eq:gameSIC} \eeq

The following result can be shown to hold.

\noindent
 {\bf Proposition 3:}
{\em Let $\bJ_k$ be a matrix having as columns the vectors in the
set
$$
\begin{array}{ccc}
\left\{ \sqrt{p_i} \bh_{i,1} \right\}_{i=1, \ldots, K} \bigcup

\left\{\sqrt{p_i}  \bh_{i,j} \right\}_{i\geq k \, ; j=-2, -1, 0}
\end{array}
$$
and define $\bM_k=\left(\bJ_k \bJ_k^T + \bM\right)$; The
non-cooperative game defined in (\ref{eq:gameSIC}) admits a unique
Nash equilibrium point $(p_k^*, \bx_k^*)$, for $k=1, \ldots, K$,
wherein
\begin{itemize}
\item[-]
$\bx^*_k=\sqrt{p_k} (\bO_k^T \bM_k \bO_k)^{-1} \bO_k^T\bh_{k,0}$
is the unique $k$-th user receive filter\footnote{Uniqueness here
means up to a positive scaling factor.} that maximizes the SINR
$\gamma_k$ given in (\ref{eq:gammaSIC2}). Denote
$\gamma_k^*=\max_{\bx_k}\gamma_k$.
\item[-]
$p_k^*=\min \{\bar{p}_k, P_{k, \max} \}$, with $\bar{p}_k$ the
$k$-th user's transmit power such that the $k$-th user's maximum
SINR $\gamma_k^*$ equals $\bar{\gamma}$, i.e. the unique solution
of the equation $f(\gamma)=\gamma f'(\gamma)$, with $f'(\gamma)$
the derivative of $f(\gamma)$.
\end{itemize}}

\noindent {\bf Proof:} The proof is omitted  due to lack of
space. \hfill
\rule{2.5mm}{2.5mm}

\subsection{Power control and unconstrained receiver design}
Finally, we consider the case in which no constraint is imposed on
the receive filter, so that the $k$-th user's SINR is written as
in (\ref{eq:gammaSIC}), and the decision rule is given by
(\ref{eq:decruleSIC}). We now consider the following maximization
\beq \ds \max_{\S_k} u_k = \max_{p_k, \bd_k} u_k(p_k, \bd_k )=
\max_{p_k}  \frac{f\left(\ds \max_{\bd_k}\gamma_k(p_k,
\bd_k)\right)}{p_k} \; , \label{eq:game5} \eeq $\forall k=1,
\ldots, K$. Now, denoting by $\bar{\bd}_k$ the maximizer of
$\gamma_k$, it is easy to show that \beq \bar{\bd}_k=\sqrt{p_k}
\bM_k^{-1} \bh_{k,0} \; ; \label{eq:mmsereceiverSIC} \eeq let us
denote by $\bar{\gamma}_k(p_k)$ the $k$-th user's SINR as
$\bd_k=\bar{\bd}_k$. Maximizing the utility with respect to the
transmit power requires instead solving the equation \beq
f(\bar{\gamma}_k(p_k))= f'(\bar{\gamma}_k(p_k))
\bar{\gamma}_k'(p_k) p_k \; . \label{eq:casinoSIC} \eeq Now,
(\ref{eq:casinoSIC}) is formally equivalent to (\ref{eq:casino})
and is quite complicated to manage. Accordingly, the same
considerations of Section III.C apply here as well, and, supported
by extensive computer simulations, we conjecture the existence of
a unique Nash equilibrium. We thus have the following

\noindent {\bf Conjecture 2:} {\em The non-cooperative game
defined in (\ref{eq:game5}) admits a unique Nash equilibrium point
$(p_k^*, {\bd}^*_k)$, for $k=1, \ldots, K$, wherein
\begin{itemize}
\item[-]
$\bd^*_k$ is given by Eq. (\ref{eq:mmsereceiverSIC}), which
maximizes the user $k$ SINR $\gamma_k$ in (\ref{eq:gammaSIC}).
Denote $\bar{\gamma}_k=\max_{\bd_k}\gamma_k$.
\item[-]
$p_k^*=\min \{\bar{p}_k, P_{k, \max} \}$, with $\bar{p}_k$ the
unique solution of Eq. (\ref{eq:casinoSIC}).
\end{itemize}}
\noindent
Also in this case, the equilibrium can be reached
through an iterative procedure. For a given set of users' transmit
powers, the receiver filter coefficients can be set equal to the
receiver in (\ref{eq:mmsereceiverSIC}); each user can then tune
its power to $p_k^*$, and these steps are  repeated until
convergence is reached.

\section{Numerical Results}

We consider now an uplink DS/CDMA system with processing gain
$N=7$, and assume that the packet length is $B=120$. Users may
have random positions with a distance from the AP
ranging from 10m to 500m. The channel impulse response $c_k(t)$
for the generic $k$-th user is assumed to be equal to $ c_k(t)=\ds
\sum_{\ell=1}^3 \alpha_{k,\ell} \delta(t- \tau_{k,\ell}) \; ,
\quad \forall k=1, \ldots, K $ with $\tau_{k,\ell}$ such that
$\tau_k + \tau_{k,\ell}$ is uniformly distributed in $[0, T_b]$
and $\alpha_{k,\ell}$ is a Rayleigh distributed random variate
with mean equal to $d_k^{-2}i_\ell$, with $d_k$ being the distance
of user $k$ from the AP, and $[i_1,\,  i_2, \,i_3]=[0.5,
\, 0.3, \, 0.2]$ . For the thermal noise level, we take
$\N=10^{-9}$W/Hz, while the maximum allowed power $P_{k,\max}$ is
$25$dB. We present here results of averaging over
$5000$ independent realizations for the users locations,
fading channel coefficients and set of spreading codes.

Figs. 1 - 2 report the achieved average utility (measured in
bits/Joule) and the average user transmit power
for the proposed non-cooperative games. As expected, the
power control game with matched filter at the receiver is the one
with the poorest performance, while the best performance is
attained by the non-linear decision-feedback receivers. It is seen
that for $K>N$ the average utility achieved by the non-linear
receivers is twice  the average utility achieved by the linear
receivers. Moreover,
constrained receivers are outperformed by unconstrained
receivers, even though the gap is not that large.

Fig. 3 reports the average fraction of users that transmit at the
maximum available power, i.e. the probability that a user
implementing a certain game is not able to achieve its target
SINR and ends up transmitting at its maximum power. As expected,
it is seen that the larger fraction corresponds to the use of a
matched filter at the receiver, while using non-linear decision
feedback receivers permits minimizing this fraction, which,
moreover, increases as the network load (i.e. number of users)
increases.

\section{Conclusions}\label{sec:Conclusions}
In this paper we have considered the problem of utility
maximization in a wireless data network through the use of a
game-theoretic approach.
The cross-layer issue of multiuser receiver design and power
control for utility maximization has been considered for the
practical scenario of an asynchronous, bandlimited and multipath
distorted CDMA system.
The case in which a non-linear decision feedback detector is adopted
has been considered. First we have derived the non-linear decision
feedback receiver maximizing the utility for each user; then, we
have  shown how the use of a non-linear multiuser receiver
provides significant performance gains, especially in the case in which the
number of users is close to or larger than the system processing
gain.
Overall, it can be stated that game theory is an attractive
mathematical tool that can be effectively used for the design of
utility-maximizing resource allocation algorithms in wireless
networks operating in practical scenarios.

\bibliographystyle{IEEEtran}
\nocite{*}
\bibliography{IEEEabrv,paper}


%

\begin{figure}[!tb]
\centerline{\hbox{\includegraphics[height=5.5cm]{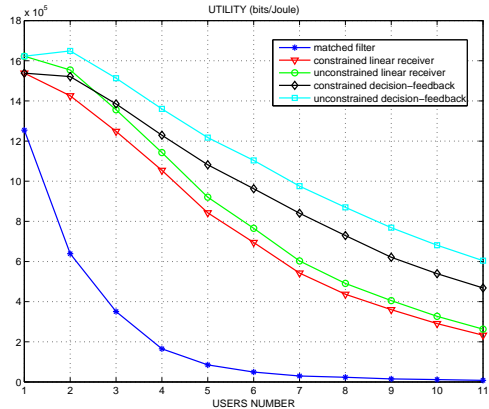}}}
\caption{Achieved average utility versus number of active users
for the proposed noncooperative games.} \label{fig:1}
\end{figure}

\begin{figure}[!tb]
\centerline{\hbox{\includegraphics[height=5.5cm]{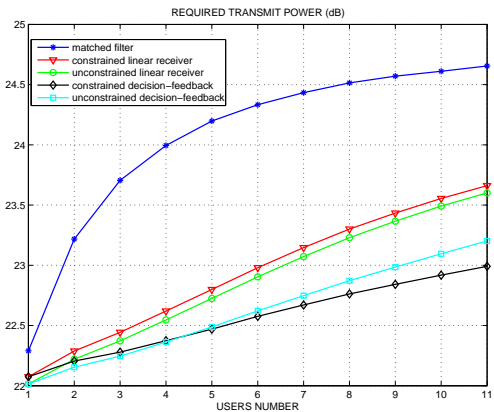}}}
\caption{Average transmit power versus number of active users for
the proposed noncooperative games.} \label{fig:2}
\end{figure}

\begin{figure}[!tb]
\centerline{\hbox{\includegraphics[height=5.5cm]{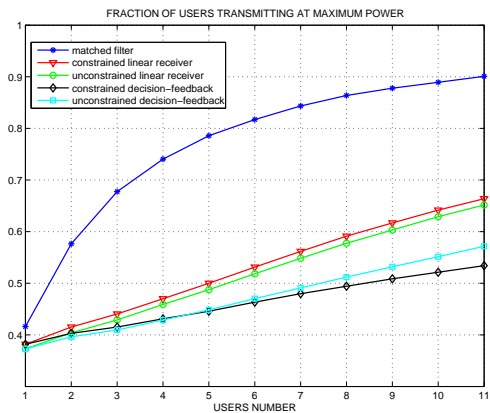}}}
\caption{Average fraction of users transmitting at the maximum
power versus number of active users for the proposed
noncooperative games.} \label{fig:3}
\end{figure}

%
%
%

\end{document}